\documentclass[11pt]{article}

\usepackage[margin=1in]{geometry}
\usepackage[T1]{fontenc}
\usepackage{lmodern}
\usepackage{microtype}
\usepackage{booktabs}
\usepackage{amsmath}
\usepackage{amssymb}
\usepackage{graphicx}
\usepackage{float}
\usepackage{placeins}
\usepackage{array}
\usepackage{xcolor}
\usepackage[numbers,sort&compress]{natbib}
\usepackage[hidelinks]{hyperref}
\usepackage{caption}
\newcommand{\tablebodyfont}{\small}
\newcommand{\tablenotefont}{\footnotesize}


\newcommand{\best}[1]{\textbf{#1}}

\title{\textbf{Qwen-MusicAVQA-7B: A Multimodal Model \\
for Music Audio-Visual QA}}

\author{
  Maryam Dehdashti\thanks{Correspondence: \texttt{dehdashti@inferencematter.ai}} \\
  Inference Matter Labs
}

\date{}

\begin{document}
\maketitle

\begin{abstract}
A common approach to adding audio to a vision-language model is to train or adapt a large
omni-modal system. We show that a lightweight alternative can be highly effective for music
audio-visual question answering (AVQA). We call this system \textbf{Qwen-MusicAVQA-7B}: a frozen
Whisper encoder connected to Qwen2-VL-7B-Instruct through learned linear projections. The \emph{same}
frozen encoder processes both the video's music track and a TTS-spoken question through separate
projectors, while the language model fuses visual frames, music, and question audio through
pretrained self-attention, with no task-specific fusion network.

On MUSIC-AVQA, our system reaches \best{96.0\%}\,$\pm$\,3.9\% accuracy across three independent
training seeds on the 7{,}402-question available-video test subset. Our central finding is that
downstream accuracy tracks how much fine-grained local temporal information the audio representation
preserves. In a matched 32-token comparison, a stride-pooled Whisper frame sequence outperforms a
globally pooled PANNs representation expanded to the same budget by 26 percentage points, even though
PANNs sees at least as much audio and uses a far larger projector. The effect is not simply sequence
versus vector: within Whisper alone, reducing temporal resolution at a fixed token budget costs a
comparable amount.

Under matched data and inputs, fine-tuned Qwen2.5-Omni-7B reaches 80.9\%, against 95.9\% for our
30\,s variant; because the systems differ in backbone and adaptation, this is a system-level
comparison. Accuracy remains high on sampled head and tail splits of the rephrased MUSIC-AVQA-R
benchmark (96.5\% and 95.6\%). Because both encoders stay frozen and the music features are cached,
the entire adaptation is cheap to train: the complete two-stage AVQA run takes approximately
5 hours on a single A100 80GB, and every run reported here fits on that one GPU.
Code is available on
\href{https://github.com/MKDehdashti/Qwen2-vl-audio}{GitHub} and checkpoints in our
\href{https://huggingface.co/MayaKD/qwen2-vl-audio}{Hugging Face repository}.
\end{abstract}

\clearpage
\section{Introduction}

Vision-language models (VLMs) reason fluently over images, video, and text, but extending them
to audio is still an active problem. A pragmatic question is whether an existing VLM backbone can be
given audio understanding through modular additions, namely an encoder and a projection layer, rather than
retraining a multimodal model from scratch. We answer this affirmatively. Our main observation is that the
amount of fine-grained local temporal information preserved in the encoder's output tracks downstream
accuracy more strongly than the capacity of the projector that consumes it.

We study music audio-visual question answering (AVQA) on the MUSIC-AVQA benchmark~\citep{li2022musicavqa},
where a system must answer closed-form questions (e.g., \emph{``how many instruments are playing?''})
about videos of live instrumental performances. The task is genuinely multimodal: it mixes visual,
auditory, and cross-modal questions, and many ostensibly ``visual'' questions (counting which instruments
\emph{sound}) cannot be answered without audio. A recent systematic analysis~\citep{you2025musicavqa}
argues that general-purpose multimodal LLMs are insufficient
for Music AVQA and highlights specialized input processing,
spatial-temporal architectures, and music-specific modeling.
Our results support the broader need for task-aware design
while identifying the audio representation as another
effective specialization lever.

Our system, which we refer to as \textbf{Qwen-MusicAVQA-7B}, grafts a pretrained Whisper
encoder~\citep{radford2023whisper} onto
Qwen2-VL-7B-Instruct~\citep{wang2024qwen2vl} through learned linear projections. Critically, the same
Whisper encoder serves two roles: it encodes the music track, and it encodes the user's question, which
we pose as synthesized speech rather than text. Spoken questions keep the interface audio-native and,
when we later compare against omni-modal models, force both systems to decode the question from audio
rather than read it as text (Section~\ref{sec:tts}). The language model then fuses visual frames, the music
track, and the spoken question entirely through its pretrained self-attention; there is no cross-modal
attention module and no learned fusion network. This contrasts sharply with prior AVQA systems, which
combine modality-specific encoders through explicit attention-based fusion~\citep{li2022musicavqa}.

A central design question is how the music encoder represents audio. A dedicated
audio-classification encoder (PANNs CNN14~\citep{kong2020panns}), used through its clip-level output,
contributes one globally pooled feature vector, whereas Whisper produces a \emph{sequence} of
frame-level representations. The
strongest-performing configurations retain time-resolved local features at ${\sim}0.94$\,s per token;
compressing up to 60\,s of audio to ${\sim}1.875$\,s per token within the same encoder is associated
with 25--27 points lower accuracy;
and the PANNs configurations, which collapse the clip into one vector and expose no explicit local
temporal sequence to the projector, reach only 66.9--69.9\%. This
pattern suggests that retaining fine-grained local temporal information is an important factor here.

Architecturally, we deliberately keep to the standard encoder--projector recipe: a frozen pretrained
encoder, a linear projector as the entire learned interface, and no new fusion mechanism. The
contribution is not a new architecture but a controlled empirical characterization---holding the
backbone, training data, and LLM token budget fixed---of which properties of the audio representation
track downstream accuracy, together with a duration-matched fine-tuned omni-modal baseline.

\medskip
\noindent Our contributions are:
\begin{enumerate}
  \item \textbf{Accuracy tracks preserved local temporal detail.} Holding the LLM token
  budget fixed at 32, a speech-trained Whisper encoder whose stride-pooled local frames are preserved
  outperforms a PANNs configuration consuming the single pooled clip-level vector by +26 points
  (Section~\ref{sec:why}), while using a far smaller projector. Within Whisper, the 32-token
  compressed variant trails both the 30\,s, 32-token variant and the full-duration chunked
  variant by 25--27 points. At the same nominal temporal resolution, the 30\,s and chunked
  configurations differ by 1.4 points, although this comparison changes duration and token count
  together. Because these comparisons change either duration or token count
  alongside resolution, they support a graded association between preserved local detail and accuracy
  rather than an isolated resolution effect; that a variant which is still a 32-token frame sequence
  collapses to 70.5\% also shows the effect is not captured by a sequence-versus-vector distinction
  alone. A speech-trained encoder with no music-specific
  pretraining suffices to reach our best accuracy; we do not evaluate a music-domain encoder
  (Section~\ref{sec:limitations}).
  \item \textbf{Fusion through pretrained LLM self-attention.} With all attention weights frozen,
  training only the projector that aligns the new modality already recovers nearly all of the final
  accuracy, and the subsequent LoRA stage shows no reliable improvement in the reported single-seed
  comparisons (Section~\ref{sec:stages}). No task-specific
  cross-modal fusion module is required.
  \item \textbf{Controlled comparison with an omni-modal baseline.} Under matched data, frames,
  question audio, and audio duration, our system exceeds a fine-tuned Qwen2.5-Omni-7B by 15 points,
  with the largest gaps on comparative and audio-visual localization question types
  (Section~\ref{sec:omni}). The systems also differ in backbone and adaptation, so we present this as a
  system-level comparison.
\end{enumerate}

\section{Related Work}

\paragraph{Audio-augmented LLMs.}
A growing line of work extends language models with audio perception. Qwen-Audio~\citep{chu2023qwenaudio}
initialized its audio encoder from Whisper-large-v2 and jointly trained it with a language model across
speech, natural-sound, music, and song tasks.
Qwen2-Audio~\citep{chu2024qwen2audio} simplified the conditioning format and scaled training data.
Qwen2.5-Omni~\citep{qwen2025omni} uses the Qwen2-Audio encoder, initialized from Whisper-large-v3 and
modified for block-wise streaming attention, within a Thinker--Talker architecture for streaming speech
generation. FAVOR~\citep{sun2024favor} introduced a causal Q-Former to capture fine-grained temporal
audio-visual relations at the frame level. video-SALMONN~\citep{sun2024videosalmonn} subsequently used a
multi-resolution causal Q-Former to connect pretrained audio-visual encoders to an LLM while retaining
fine-grained temporal information across visual frames, speech, audio events, and music.
Our approach shares Qwen-Audio's use of a Whisper-initialized audio encoder with an LLM, but keeps
that encoder frozen, applies it to a VLM backbone (Qwen2-VL), and adds a second audio path for music.
Unlike FAVOR and video-SALMONN, we use no learned Q-Former or task-specific fusion module; the
separate linear projectors only place each stream in the LLM embedding space.

\paragraph{Audio-visual question answering.}
MUSIC-AVQA~\citep{li2022musicavqa} pairs instrumental-performance videos with closed-form questions
spanning audio, visual, and audio-visual reasoning. The original system encodes audio with
VGGish~\citep{hershey2017vggish}, visual frames with a ResNet-18, and questions with an LSTM over
projected word embeddings, then
fuses them with spatial and temporal cross-modal attention modules. Subsequent work has largely retained
the encoder-plus-explicit-fusion paradigm: LAVisH~\citep{lin2023lavish} adapts frozen vision transformers
with lightweight adapters for cross-modal learning; DG-SCT~\citep{duan2023dgsct} uses dual-guided
spatial-channel-temporal attention. Amuse~\citep{diao2024amuse} adds interactive multimodal encoders,
explicit rhythm and source predictors aligned to time, and music-region features, reporting 83.52\%
overall accuracy on the full MUSIC-AVQA test split. Sparsify~\citep{diao2025sparsify} introduces sparse
masking and token merging to reduce representational redundancy, reporting 81.75\% overall accuracy while
also reducing training cost. \citet{ma2024musicavqar} introduce MUSIC-AVQA-R, a robustness benchmark that
rephrases the test questions to expose methods that exploit surface-level question patterns rather than
reasoning over the audio and video; we evaluate our final model on it
(Section~\ref{sec:robustness}) to test its robustness to
rephrased questions.
Meerkat~\citep{chowdhury2024meerkat} tackles fine-grained audio-visual grounding in space and time
using optimal-transport alignment, though it targets general audio-visual scenes rather than music
specifically. These systems generally rely on explicit cross-modal interaction, alignment, or fusion
components. We dispense with
bespoke fusion entirely: all three modalities are routed through a single language model whose
pretrained self-attention performs fusion implicitly; this simplicity is what makes the
representation comparisons in Section~\ref{sec:why} interpretable.

\paragraph{Audio representations for understanding.}
How to represent audio for a downstream reasoner is contested. PANNs~\citep{kong2020panns} are CNNs
trained on AudioSet for tagging; while CNN14 computes intermediate feature maps over time, its
clip-level output is a single globally pooled classification vector, which is what we consume. Sequence
encoders such as Whisper~\citep{radford2023whisper} yield frame-level representations along time. A recurring design
choice is whether to pool audio to a compact vector or to preserve its temporal sequence; recent
large-scale contrastive work likewise spans speech, music, and sound to learn broadly transferable
audio representations~\citep{vyas2025semanticaudio}. Concurrent work adapts Whisper to non-speech
domains: Whisper-AuT~\citep{qiu2026whisperaut} reports gains over Whisper on non-speech tasks including
GTZAN music-genre classification, while UniWhisper~\citep{uniwhisper2026} improves average probe
performance across a broader set of speech, environmental-sound, and music tasks. Both are statements
about per-frame feature \emph{quality}. Our observation is complementary:
for music question \emph{answering}, an unmodified speech-trained encoder is a strong music encoder once
its sequence of local frames is preserved, even though we do not independently measure per-frame feature
quality in our setting.

\paragraph{Music and frame-wise models.}
MERT~\citep{li2023mert} uses music-oriented self-supervised acoustic and musical teachers and reports
strong transfer across 14 music-understanding tasks. MusiLingo~\citep{deng2023musilingo} is a close
modular precedent, aligning a frozen MERT encoder with a frozen language model through a single projection
layer for music captioning and query response, though it pairs one music encoder with the LLM and never
contrasts local against pooled representations. FLAM~\citep{wu2025flam} trains a frame-wise
audio--language objective for open-vocabulary event localization, directly targeting the loss of local
event information in globally summarized audio, and MusTBENCH~\citep{kwon2026mustbench} finds that current
music language models struggle to ground answers in the correct temporal regions. This frame-wise line
makes the case that local, temporally resolved audio matters, but it argues so through a
representation-learning objective and a grounding benchmark rather than by isolating that property inside a
downstream QA system with the reasoner held fixed.

\begin{table}[H]
\centering
\setlength{\tabcolsep}{4pt}
\renewcommand{\arraystretch}{1.25}
\caption{\textbf{Representative architectural approaches to music-language and audio-visual
reasoning.} A qualitative positioning summary across different tasks and datasets, not a performance
comparison: accuracy is omitted by design (see Table~\ref{tab:main} and Section~\ref{sec:omni} for the
controlled quantitative comparisons). The systems shown are a representative sample rather than an
exhaustive list. ``Q'' denotes the question.}
\label{tab:arch}
\tablebodyfont
\makebox[\textwidth][c]{%
\begin{tabular}{>{\raggedright\arraybackslash}p{0.148\textwidth}
                >{\raggedright\arraybackslash}p{0.195\textwidth}
                >{\raggedright\arraybackslash}p{0.175\textwidth}
                >{\raggedright\arraybackslash}p{0.195\textwidth}
                >{\raggedright\arraybackslash}p{0.202\textwidth}}
\toprule
System & Inputs $\rightarrow$ output & Audio representation & Connector / fusion & Main architectural distinction \\
\midrule
AVST~\citep{li2022musicavqa}          & audio+video, text Q $\rightarrow$ answer label      & VGGish audio features             & task-specific spatial-temporal AV attention & conventional encoder-fusion AVQA \\
LAVisH~\citep{lin2023lavish}          & audio+video, text Q $\rightarrow$ answer label      & frozen ViT audio features         & lightweight cross-modal adapters            & parameter-efficient adapter adaptation \\
MusiLingo~\citep{deng2023musilingo}   & music, text Q $\rightarrow$ generated text          & frozen MERT music features       & single projection to a frozen LLM           & modular music-to-language alignment \\
video-\allowbreak SALMONN~\citep{sun2024videosalmonn} & video+audio+\allowbreak speech, text instr.\ $\rightarrow$ generated text & audio-visual frame-feature sequence & multi-resolution causal Q-Former        & specialized temporal connector \\
Qwen2.5-Omni~\citep{qwen2025omni}$^\dagger$ & text / audio / image / video $\rightarrow$ generated text & Qwen2-Audio-derived feature sequence & integrated omni-modal architecture          & large native omni-modal system \\
\midrule
\textbf{Ours}                         & video+music, spoken Q $\rightarrow$ generated text answer & local Whisper frame-feature sequence & separate linear projectors; LLM self-attention & shared frozen encoder for music and query; no task-specific fusion \\
\bottomrule
\end{tabular}%
}

\vspace{2pt}
\raggedright\tablenotefont
$^\dagger$ The only row we evaluate quantitatively: Qwen2.5-Omni is fine-tuned by us under matched data,
inputs, and Stage-2 hyperparameters (Table~\ref{tab:main}, Section~\ref{sec:omni}). We evaluate only its
generated-text answer and do not use or evaluate its native speech generation. The other prior systems are
cited for architectural context only.
\end{table}

\FloatBarrier

\paragraph{This work.}
Prior AVQA work has largely fixed the audio encoder and focused on explicit cross-modal
fusion, while omni-modal systems emphasize broader audio pretraining and scale. Recent
music-language and audio-visual models also introduce specialized connectors to preserve
fine-grained temporal information. What remains less clear is how the form of the audio
representation itself affects downstream reasoning when the language model and fusion
mechanism are shared. We study this question within a common Qwen2-VL framework by
comparing globally pooled audio representations with sequences of distinct local features,
while allowing each representation its corresponding projection layer. Table~\ref{tab:arch} contrasts
this design with representative prior systems.

\clearpage

\section{Method}

\subsection{Overview: shared Whisper encoder with dual audio paths}
\label{sec:arch}

\begin{figure}[p]
\centering
\makebox[\textwidth][c]{\includegraphics[width=1.04\textwidth]{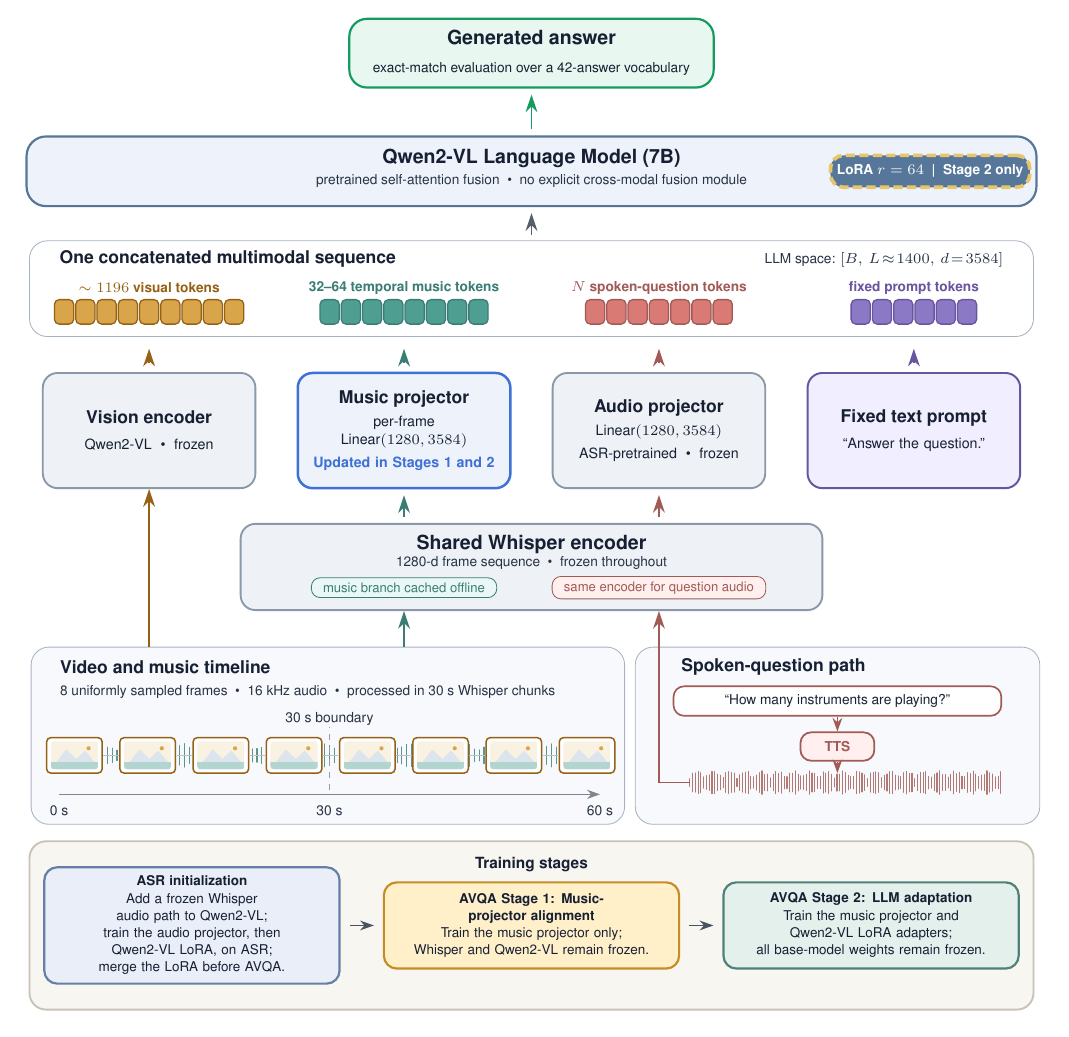}}
\caption{\textbf{ASR initialization followed by two explicit AVQA training stages.}
Before AVQA training, speech audio is encoded by frozen Whisper while the audio projector and an ASR
LoRA adapter align the spoken-question pathway with Qwen2-VL; the ASR adapter is then merged, and these
weights initialize every AVQA run. In AVQA Stage~1, visual, music, spoken-question, and fixed-prompt
representations are concatenated into one token sequence, but only the music projector is updated; the
Whisper encoder, vision encoder, question-audio projector, and Qwen2-VL base weights remain frozen. In
Stage~2, the same four token groups are regenerated and concatenated, the music projector remains
trainable, and new Qwen2-VL LoRA adapters are updated while all base-model weights remain frozen. Music is
processed in 30-second chunks, each pooled separately to 32 tokens, so temporal resolution remains
approximately constant while the number of music tokens scales with clip duration. Qwen2-VL performs
multimodal fusion through its pretrained self-attention without a task-specific fusion network.}
\label{fig:arch}
\end{figure}

Figure~\ref{fig:arch} shows the architecture. We extend Qwen2-VL-7B-Instruct with two audio paths that inject
tokens into the LLM's embedding space, alongside the inherited visual path: a \emph{question path} that
encodes the question, posed as synthesized speech, through a projector first aligned by ASR
pretraining (Section~\ref{sec:asr}), and a \emph{music path} that encodes the video's music track. Both paths use the same frozen Whisper-large-v3-turbo
encoder~\citep{radford2023whisper} (1280-dim frame outputs), each with its own linear projector into
the LLM's 3584-dim hidden space. The projectors are kept separate so the
speech and music representations can be aligned independently with the LLM space, while the same
frozen Whisper encoder is used by default for both streams, reusing one pretrained feature family
rather than deploying a separate music-specialist backbone. Section~\ref{sec:main} finds that
preserving local frame-level features from this shared encoder already reaches 97.3\% accuracy on
MUSIC-AVQA, leaving limited aggregate-accuracy headroom for a dedicated music encoder
(Section~\ref{sec:limitations}). The system is built in three steps, described in order below: ASR
pretraining aligns the question path with the LLM (Section~\ref{sec:asr}); the music path is added and
its projector aligned on AVQA data (Sections~\ref{sec:music} and~\ref{sec:training}); and a light LoRA
stage adapts the LLM (Section~\ref{sec:training}).

\begin{samepage}
\paragraph{Implicit fusion.}
Visual, music, and question tokens are concatenated into one sequence and processed by the LLM's
self-attention. There is no cross-modal attention module: fusion is whatever the pretrained attention
learns to do over the concatenated tokens. This is the central architectural simplification relative to
prior AVQA systems. Concretely, each example is a single chat turn containing, in order, the video
frames, the music clip, the spoken question, and the fixed text instruction ``Answer the question.'';
the model generates the answer as free text.
\end{samepage}

\subsection{ASR pretraining}
\label{sec:asr}

Qwen2-VL is a vision-language model: out of the box it has no audio pathway, so before it can hear a
music track or a spoken question it must first learn to read the Whisper encoder's output. Supplying
that ability is the purpose of ASR pretraining. The audio projector, the single linear layer bridging
Whisper's 1280-dim frames to the LLM's 3584-dim token space, is randomly initialized, and to Qwen2-VL
its outputs are meaningless. Speech recognition is a natural task to align it: transcribing speech
forces the projector to map audio frames into the region of token space where the LLM already expects
the corresponding words. Without this step, AVQA training would have to learn audio alignment from
scratch on a smaller, noisier signal; instead, AVQA training always begins from an already-aligned
audio path.

Concretely, we freeze both large models (the Qwen2-VL LLM and the Whisper encoder) and train only the
linear audio projector on a 20{,}000-utterance subset (clips $\leq$15\,s) of the LargeScaleASR corpus
(\texttt{speechbrain/\allowbreak LargeScaleASR}, \emph{small} configuration); a second stage then adds
LoRA~\citep{hu2022lora} ($r{=}64$) on Qwen2-VL's attention layers, on the same data. The resulting
model reaches 4.85\% word error rate on LibriSpeech test-clean (full split, $n{=}2{,}620$,
normalized), confirming that the projector has learned a mapping the LLM can decode as text. We do not
treat ASR as a contribution and do not tune for it. For reference, Qwen2.5-Omni reports 1.8\% WER on
LibriSpeech test-clean~\citep{qwen2025omni}; recognition quality is not our objective.
These ASR weights, with the LoRA adapters merged into the base model, initialize every one of our
AVQA experiments.

\subsection{Question path: TTS-spoken questions through the ASR-aligned encoder}
\label{sec:tts}

We pose each question as synthesized speech and feed it through the Whisper encoder rather than
tokenizing it as text, which keeps the question interface audio-native. Questions are synthesized with a neural TTS
system (\texttt{edge-tts}, voice \texttt{en-US-AriaNeural}), once per unique question string
(${\sim}2{,}800$ unique strings cover all ${\sim}46$K questions), and the waveforms are cached.

At training and inference time, the cached waveform is resampled to 16\,kHz by the processor and
encoded by the frozen Whisper encoder in the forward pass; unlike the music path below, the
question's Whisper features are not precomputed. They could be cached the same way, but questions
are only a few seconds long and the Whisper encoder is already in memory, so encoding them live
costs little. The encoder output is a frame sequence that the
ASR-aligned audio projector ($\mathrm{Linear}(1280, 3584)$) maps into the LLM's hidden space. In the
chat-formatted input the question occupies a single \texttt{<|audio\_pad|>} placeholder token, which
the processor expands to one token per encoder frame (${\sim}50$ frames per second of speech), so the
number of question tokens varies with the duration of the synthesized question.

\subsection{Music path}
\label{sec:music}

The music path is precomputed offline. For each video, ffmpeg extracts the audio track as a mono
waveform resampled to 16\,kHz, Whisper's input rate (the alternative encoder of
Section~\ref{sec:why}, PANNs, uses its native 32\,kHz rate). The frozen Whisper encoder is then run over
the waveform in 30\,s windows, the maximum its learned positional embeddings support; each window
yields 1{,}500 frames, which we \emph{stride-pool}, i.e.\ average-pool along time into 32 equal-width
bins (implemented as adaptive average pooling), giving 32 frames per 30\,s chunk: a nominal
${\sim}0.94$\,s/token for complete chunks. A trailing chunk shorter than 5\,s is discarded, and a
retained partial chunk is pooled to 32 tokens as well, so its tokens each cover proportionally less
time; the ${\leq}61$\,s clips in this dataset therefore yield one or two chunks, i.e.\ 32 or 64 music
tokens. The pooled features are cached to disk, so the music encoder is never part of the training
graph.

The \emph{music projector} maps the cached features to music tokens. For
Whisper-as-music-encoder the projector maps each pooled frame to one token
($\mathrm{Linear}(1280, 3584)$), preserving a temporal sequence of 32--64 tokens depending on clip
duration (Section~\ref{sec:why}). For the single-vector PANNs variants
(Section~\ref{sec:why}), the projector instead expands one global vector into a fixed number of
tokens ($\mathrm{Linear}(D, n\!\cdot\!3584)$). In the input, the music clip is delimited by three
special tokens added to the vocabulary, \texttt{<|music\_start|>}, \texttt{<|music\_pad|>}, and
\texttt{<|music\_end|>}; the processor expands the single \texttt{<|music\_pad|>} placeholder to the
configured number of music tokens, mirroring the question path's mechanism.

The music encoder need not be Whisper, and the Whisper features need not span only 30\,s. To study how
the form of the music representation affects accuracy, we also evaluate PANNs and Whisper
configurations that vary clip duration and temporal resolution; these variants are defined alongside
their results in Section~\ref{sec:why}.

The PANNs and Whisper comparisons are matched on the downstream token budget where stated, but not
always on input duration. PANNs CNN14 is run over the \emph{entire} extracted waveform (typically
30--60\,s) and pools it to one 2048-d vector, so it sees at least as much audio as any Whisper variant.

Precomputing the music path offline serves two purposes. First, the music encoders are frozen, so
their features never change across epochs, and caching avoids re-encoding 30--60\,s clips on every
forward pass. Second, the ablations of Section~\ref{sec:why} swap the music encoder for PANNs; with
features cached to disk, the encoder does not need to be loaded during training.

\subsection{Training}
\label{sec:training}

AVQA training begins from the merged ASR weights with a randomly initialized
music projector. \textbf{Stage 1} freezes the LLM, the Whisper encoder, and the audio projector, and
trains only the music projector, the alignment step that teaches the model to read music.
\textbf{Stage 2} continues from the Stage-1 checkpoint and adds LoRA on the LLM attention layers. Our default configuration freezes the
question-audio projector during Stage~2 as a conservative choice that leaves the ASR-aligned question
path unchanged; Section~\ref{sec:stages} reports the corresponding single-seed comparison.

\FloatBarrier

\section{Experiments}

\subsection{Setup}
\label{sec:setup}

MUSIC-AVQA~\citep{li2022musicavqa} is a benchmark for reasoning over musical performances: short
videos, collected from YouTube, of people playing instruments, both solo and in multi-instrument
ensembles, paired with roughly 46K closed-form questions. Each example is a \emph{video--question pair}
whose question requires perceiving what is heard, what is seen, or both jointly, so that many cannot be
answered from a single modality alone (for instance, counting how many instruments are
\emph{sounding}); answers come from a fixed 42-answer vocabulary. Questions span nine types along two
axes (modality: Audio, Visual, Audio-Visual; reasoning: Existential, Counting, Location, Comparative,
Temporal). Appendix~\ref{app:examples} summarizes how these axes form the nine official question types
and provides one representative example of each.

About 20\% of the benchmark's videos were no longer downloadable when we ran our experiments, so we work
throughout with the available-video subset: we train on 8{,}000 pairs (the first 8{,}000 in the
dataset's fixed order, no random subsampling) and evaluate on a fixed 7{,}402-pair test set. Every
model we run---including all ablations and both Omni baselines---is evaluated on this same
7{,}402-pair subset. All fine-tuned models we run use the same 8{,}000-pair training subset. Because our
model and configuration choices were developed on it, absolute accuracy may be optimistic relative to a
fresh held-out benchmark, though the relative comparisons among our variants (identical protocol
throughout) are more reliable. The official splits share source videos across train and test, so
evaluation measures held-out video--question pairs rather than generalization to unseen videos; this
is a property of the benchmark's official splits and applies equally to the prior work in
Table~\ref{tab:main}, which uses those same splits.

The model answers by greedy autoregressive generation rather than through a 42-way classification
head, for two reasons: such a head is exactly the kind of task-specific module our design avoids, and
free-form generation keeps the output interface identical to Qwen2.5-Omni, the baseline we evaluate
under matched inputs, so that comparison is not confounded by the answering
mechanism. Generation is capped at 5 tokens, and we report exact-match accuracy against the fixed
42-answer vocabulary after lowercasing and punctuation stripping, on a consistent evaluation set across
all variants.

To assess sensitivity to initialization, we retrain the full pipeline (Stage 1 and Stage 2) under three
random seeds, \{42, 1234, 2026\}. Seed 42 is the fixed default for every ablation, encoder variant, and
the Omni baseline, so all per-type comparisons are made at a single controlled seed; we report and
analyze the three-seed spread in Section~\ref{sec:seeds}.

\paragraph{Implementation.}
All experiments use a single NVIDIA A100 80GB and HuggingFace Transformers with the TRL supervised
fine-tuning trainer, PEFT LoRA adapters, fused AdamW, and a linear-decay learning-rate schedule.
Stage~1 uses learning rate $10^{-4}$ with 50 warmup steps; Stage~2 uses $2{\times}10^{-5}$ with
50 warmup steps. We use bf16/tf32, an effective batch size of 16, one epoch per stage, and LoRA
$r{=}64,\ \alpha{=}128$ on the attention projections. Each example contains eight uniformly sampled
video frames ($\sim1196$ visual tokens), precomputed and cached alongside the music features. The
Whisper music projector is a per-frame $\mathrm{Linear}(1280,3584)$ layer with 4.6M parameters;
pooled PANNs representations instead use $\mathrm{Linear}(D,n\!\cdot\!3584)$ expansion heads. We monitor a fixed 200-example cap of the 3{,}698-example available validation split for
logging only and do not select checkpoints by validation performance.

\subsection{Main results}
\label{sec:main}

Table~\ref{tab:main} places our best model against the original benchmark, subsequent encoder-fusion
methods, and Qwen2.5-Omni-7B. Our representative model reaches \best{97.3\%}. Published results are
included for historical context only, and we deliberately do not compute differences against them: prior
systems use the full official training and test splits, whereas our experiments use 8{,}000 training
pairs, roughly a quarter of the official ${\sim}32$K, and the available-video test subset
(Section~\ref{sec:setup}), which could be easier, harder, or similar. We therefore report modality-level
breakdowns only for the systems we run ourselves, not for the published baselines.

Even given our exact inputs (frames, music, and the TTS-spoken question), zero-shot Qwen2.5-Omni scores
only 56.8\%, close to our own no-question vision floors (53.7\% for Whisper-60s-chunked and 55.8\% for
PANNs-32, Section~\ref{sec:modality}) and below all PANNs and Whisper audio variants in our study. The
modality-level breakdown locates the shortfall: zero-shot Omni reaches 83.1\% on Audio questions but
only 52.0\% on Audio-Visual and 52.1\% on Visual questions, so matched fine-tuning moves Audio accuracy by 2.0 points
while adding 25.8 on Audio-Visual and 32.8 on Visual. The zero-shot deficit is thus concentrated in
cross-modal and visual reasoning rather than in audio perception, although a zero-shot model is also
penalized by our closed-vocabulary exact-match protocol, so we do not read this as a pure perception
result. Modality access alone is therefore not enough without
task-specific adaptation. Fine-tuned under matched conditions, Omni still trails our model, a gap we
analyze in Section~\ref{sec:omni}.

\begin{table}[H]
\centering
\caption{\textbf{Main results on the available-video MUSIC-AVQA test subset ($n{=}7{,}402$).} Published
encoder-fusion baselines are included for context only. Qwen2.5-Omni is evaluated on the same
$7{,}402$-question subset and inputs as our models, with the remaining system-level differences
described in Section~\ref{sec:omni}.
``TTS + native'': the Omni model receives the same TTS-spoken question and decodes it through its own
native audio pathway. The Whisper-60s-chunked value is the seed-42 representative run; the three-seed
result is $96.0\%\pm3.9\%$ (Section~\ref{sec:seeds}). Bold marks the best value within the lower,
run-by-us block only; we make no claim against the published block.}
\label{tab:main}
\tablebodyfont
\begin{tabular}{l l c c c c}
\toprule
Model & Question enc. & Overall & Audio & AV & Visual \\
\midrule
\multicolumn{6}{l}{\textit{Published; full official train/test splits---not directly comparable}} \\
MUSIC-AVQA / AVST~\citep{li2022musicavqa} & text & 71.6$^\dagger$ & --- & --- & --- \\
DG-SCT~\citep{duan2023dgsct} & text & 74.6$^\dagger$ & --- & --- & --- \\
LAVisH~\citep{lin2023lavish} & text & 76.1$^\dagger$ & --- & --- & --- \\
Sparsify~\citep{diao2025sparsify} & text & 81.8$^\dagger$ & --- & --- & --- \\
Amuse~\citep{diao2024amuse} & text & 83.5$^\dagger$ & --- & --- & --- \\
\midrule
\multicolumn{6}{l}{\textit{Run or evaluated by us on the same $n{=}7{,}402$ test subset; fine-tuned rows use 8{,}000 training pairs}} \\
Qwen2.5-Omni-7B (zero-shot)$^\S$ & TTS + native & 56.8 & 83.1 & 52.0 & 52.1 \\
Qwen2.5-Omni-7B (fine-tuned) & TTS + native & 80.9 & 85.1 & 77.8 & 84.9 \\
Ours: Whisper-30s & TTS + Whisper & 95.9 & 95.5 & 95.1 & \best{97.6} \\
Ours: Whisper-60s-chunked & TTS + Whisper & \best{97.3} & \best{97.4} & \best{97.3} & 97.3 \\
\bottomrule
\end{tabular}

\vspace{2pt}
\raggedright\tablenotefont
$^\dagger$ AVST (71.59, shown as 71.6), DG-SCT (74.62, shown as 74.6), LAVisH (76.10,
shown as 76.1), and Sparsify (81.75, shown as 81.8) follow the comparison reported by
\citet{diao2025sparsify} on the original MUSIC-AVQA test split. The original AVST
paper~\citep{li2022musicavqa} reports 71.52\%. Amuse's 83.52\% (shown as 83.5) follows its original
paper~\citep{diao2024amuse}; that paper reports 85.16\% on the later MUSIC-AVQA v2.0 split.\\
$^\S$ $n{=}7{,}402$, audio-matched: the base omni model is given the same frames, music track, and
TTS-spoken question as the fine-tuned row, differing only in the absence of the LoRA adapter. The low score
is consistent with the absence of task-specific alignment and should not be read as the model's general
capability ceiling (see Section~\ref{sec:main}).
\end{table}

\subsection{Seed variance}
\label{sec:seeds}

Across the three training seeds \{42, 1234, 2026\} (Section~\ref{sec:setup}), the full pipeline reaches
97.3\%, 91.6\%, and 99.1\% on the test set, for a mean of \textbf{96.0\,$\pm$\,3.9\%} (sample standard
deviation; range $7.5$ points). We retain the canonical seed-42 run (97.3\%) as the representative model
for all per-type tables; for the headline configuration it is also the \emph{median} of the three
seeds, which rules out upward selection. The spread is not uniform across question types: it is
concentrated in the comparative categories, where one seed's accuracy fell to ${\sim}76\%$ while another
reached $100\%$, with the remaining types stable to within a couple of points. The ablation variants
were each evaluated with a single seed; the central effects they show (e.g.,\ the $+26$-point
Whisper-30s versus PANNs-32 gap, and ${\sim}+15$ over fine-tuned Omni at matched audio duration) are much
larger than this cross-seed spread and support the qualitative pattern, though we do not have a
per-effect uncertainty estimate. Small inter-variant gaps, however, such as the tuned and frozen
question-projector variants (95.5\% and 97.3\%, respectively), fall within this seed band and should not
be read as established orderings.

Final Stage-2 losses are nearly identical across seeds (train loss 0.211--0.217; validation loss
0.198--0.207) despite this 7.5-point accuracy range, and the two metrics do not rank consistently: seed
1234 has the lowest training loss and the second-lowest validation loss of the three, yet the lowest
accuracy (91.6\%). Aggregate loss therefore does not track the seed-to-seed accuracy variation. Because
the spread is concentrated in the comparative categories while other types remain stable (above), this is
consistent with seed-dependent shifts near a few closely scored generation decisions rather than a broad
difference in model fit, though confirming that mechanism would require examining per-token margins
rather than aggregate loss.

\subsection{Music-encoder comparison}
\label{sec:why}

Table~\ref{tab:encoders} reports our core ablation: we swap the music encoder while holding the
downstream LLM, training framework, and evaluation protocol fixed. The headline Whisper-30s versus
PANNs-32 comparison additionally matches the 32-token budget.
\textbf{PANNs CNN14}~\citep{kong2020panns} (32\,kHz input) contributes a 2048-dim clip-level embedding,
a single globally pooled vector over the whole track; \textbf{Whisper}~\citep{radford2023whisper}, the
same frozen ASR encoder (Section~\ref{sec:music}), contributes a 1280-dim frame \emph{sequence}. For Whisper we compare three
configurations that vary clip duration and temporal resolution: \textbf{Whisper-30s} (first 30\,s, 32
tokens, ${\sim}0.94$\,s/token), \textbf{Whisper-60s-chunked} (each 30\,s chunk pooled independently,
32--64 tokens, a nominal ${\sim}0.94$\,s/token), and \textbf{Whisper-60s-compressed} (the chunks
concatenated then pooled to 32 tokens, covering up to ${\sim}60$\,s at a nominal
${\sim}1.875$\,s/token).

\paragraph{Duration and resolution, within Whisper.}
Most clips are ${\sim}60$\,s but Whisper sees only 30\,s at a time. In our runs, extending to the full
clip improved accuracy only in the configuration that preserved temporal resolution. Pooling each chunk
independently (Whisper-60s-chunked)
improves overall accuracy by 1.4 points over Whisper-30s (95.9\%\,$\rightarrow$\,97.3\%). The two largest
category-level gains from full-duration audio are Audio/Comparative ($+6.8$, 88.7\%\,$\rightarrow$\,95.5\%)
and AV/Temporal ($+6.0$, 86.2\%\,$\rightarrow$\,92.2\%). The AV/Temporal gain has the clearest
duration-based interpretation, because some event-ordering questions can depend on events after the
first 30\,s; Audio/Comparative belongs to the category family with the widest
cross-seed variation (Section~\ref{sec:seeds}), so its exact gain should be read cautiously. For scale,
Visual/Location moves by 1.0 point (96.6\%\,$\rightarrow$\,95.6\%) under this audio-only change, which
gives an empirical indication of how large small single-run category fluctuations can be; the 1.4-point
overall gain should therefore be treated as descriptive, whereas the two ${\sim}6$-point category
movements are the more notable effects.

Naively compressing the same 60\,s into 32 tokens instead (Whisper-60s-compressed) yields
${\sim}1.875$\,s/token; accuracy falls to 70.5\%. The compressed variant trails the 30\,s, 32-token
variant by 25.4 points and the 60\,s chunked variant by 26.8 points. Neither comparison isolates
resolution: the first also changes duration, while the second also changes token count. At fixed
nominal resolution, extending duration while proportionally increasing the number of tokens
(Whisper-30s versus Whisper-60s-chunked) changes overall accuracy by only 1.4 points in this run.
This comparison does not separately identify duration and token-count effects or exclude their
interaction with compression. We therefore treat the compressed variant's 70.5\% result as strong
associative evidence that preserving fine-grained local detail matters, rather than as an isolated
measurement of a resolution effect. Additional resolutions and seeds would be required to
characterize the relationship.

\paragraph{Preserved local detail, across encoders.}
The same lesson holds across encoders, at the extreme. Whisper-30s outperforms PANNs-32 by
\best{+26 points} although both feed exactly 32 tokens to the same LLM. At the LLM interface, their
salient representational difference is
what the encoder emits: the PANNs configuration we consume provides one global vector that a linear
layer expands to 32 tokens,
whereas Whisper produces 32 distinct local frames, and the learned expansion cannot reconstruct local
detail that pooling discarded. Whisper's frame sequence may preserve event timing, intensity variation,
and timbre that a single pooled vector cannot express as a time-indexed sequence. Duration does not
explain the gap, and runs the other way: PANNs
pools the \emph{entire} 30--60\,s waveform (Section~\ref{sec:music}) while Whisper-30s sees only the
first 30\,s, so the pooled encoder is given at least as much audio. Capacity likewise runs the other way:
PANNs-32 expands its 2048-d vector with a ${\sim}235$M-parameter head, whereas Whisper's per-frame
projector is only ${\sim}4.6$M. Nor does the question interface: the question path
(TTS\,$\rightarrow$\,Whisper) is byte-identical in both, and only the music branch changes. (We test
absolute-score template exploitation separately in Section~\ref{sec:robustness}.) The pattern recurs
within PANNs, where expanding $8\rightarrow32$ tokens adds only +3 points: however many tokens the
learned head emits, their information content is bounded by the single pooled vector.

Read together with the within-Whisper result above, these comparisons are of similar magnitude: the
comparisons involving the compressed variant show 25--27-point gaps, and the matched Whisper--PANNs
comparison shows an approximately 26-point gap. We therefore do not frame the finding as a categorical
sequence-versus-vector distinction,
since a configuration that is still a 32-token sequence of distinct frames can fall to 70.5\%. What the
evidence supports is the graded statement that accuracy tracks how much fine-grained local temporal
information survives into the projected tokens, with global pooling as the limiting case in which no
explicit time-indexed local sequence remains.

Across these comparisons, preserved local temporal detail tracks accuracy more consistently than
projector capacity. Because PANNs and Whisper differ in architecture and pretraining, we read this as
identifying which property tracks performance rather than as an isolated causal test; we did not run
shuffled- or reversed-token controls, so we attribute the gain to preserving distinct local features and
not to temporal \emph{order} specifically.

\begin{table}[H]
\centering
\caption{\textbf{Audio-encoder ablation (MUSIC-AVQA test, $n{=}7{,}402$).} All variants share the
same downstream LLM, training stages, and evaluation protocol. PANNs pools the full clip into one vector;
Whisper preserves a frame sequence at the listed temporal resolution, quoted as the nominal value for
complete 30\,s chunks. Whisper-30s and PANNs-32 both feed
32 music tokens, while PANNs sees at least as much audio. ``Proj.'' is the music-projector parameter
count; each row is a single training run (Section~\ref{sec:seeds}). AV/Temp.\ = Audio-Visual/Temporal.}
\label{tab:encoders}
\tablebodyfont
\setlength{\tabcolsep}{4.5pt}
\begin{tabular}{l l c c c c c}
\toprule
 & & & & & \multicolumn{2}{c}{Accuracy (\%)} \\
\cmidrule(l){6-7}
Music encoder & Representation & Audio in & \# tok.\ & Proj. & Overall & AV/Temp. \\
\midrule
\multicolumn{7}{l}{\textit{Pooled: a single global vector expanded to tokens}} \\
PANNs-8 & classification, 2048-d & full clip & 8 & 59M & 66.9 & 69.9 \\
PANNs-32 & classification, 2048-d & full clip & 32 & 235M & 69.9 & 69.4 \\
\midrule
\multicolumn{7}{l}{\textit{Whisper: a 1280-d frame sequence (one token per frame)}} \\
Whisper-30s & 0.94\,s/token & first 30\,s & 32 & 4.6M & 95.9 & 86.2 \\
Whisper-60s-chunked & 0.94\,s/token & ${\sim}60$\,s & 32--64 & 4.6M & \best{97.3} & \best{92.2} \\
Whisper-60s-compressed & 1.875\,s/token & ${\sim}60$\,s & 32 & 4.6M & 70.5 & 64.7 \\
\bottomrule
\end{tabular}
\end{table}
\FloatBarrier

\subsection{Projector-only alignment recovers near-final accuracy}
\label{sec:stages}

Table~\ref{tab:stages} evaluates whether Stage~2 LLM adaptation provides a measurable benefit after the
music projector has been aligned with the frozen language model. For PANNs-8, Stage~2 changes accuracy
from 66.2\% to 66.9\%. For Whisper-60s-chunked, projector-only Stage~1 already reaches 96.0\%, while the
two Stage~2 configurations reach 95.5\% and 97.3\%. These changes are all small relative to the observed
cross-seed variation (Section~\ref{sec:seeds}) and do not establish a reliable Stage~2 improvement. The
positive result is instead that projector-only alignment with the LLM frozen already recovers nearly all
final accuracy.

The two Whisper-60s-chunked rows also isolate the question-projector choice introduced in
Section~\ref{sec:training}; both share the same seeded Stage-1 checkpoint, data, and hyperparameters
($r{=}64$ LoRA), so the frozen projector is the sole difference. The frozen-question-projector variant
is 1.8 points higher overall than the tuned variant (95.5\%~$\rightarrow$~\best{97.3\%}) and also has
higher comparative-category scores at the shared seed. However, these are single-seed differences, and
the comparative categories exhibit the widest cross-seed variation (Section~\ref{sec:seeds}). They
therefore do not establish a reliable advantage or a calibration mechanism. We retain the frozen
question projector as a conservative default because it leaves the ASR-aligned question pathway
unchanged.

\begin{table}[H]
\centering
\caption{\textbf{Projector-only Stage~1 versus Stage~2 LLM adaptation.}
Stage~1 already reaches 96.0\% with Whisper-60s-chunked while the LLM
remains frozen. The subsequent Stage~2 changes ($+0.7$, $-0.5$, and
$+1.3$ points) are descriptive single-seed differences and are small
relative to the observed cross-seed variation (Section~\ref{sec:seeds}); the table therefore
supports near-final projector-only performance, not a reliable ordering
among the Stage~2 configurations.}
\label{tab:stages}
\tablebodyfont
\begin{tabular}{l c c c}
\toprule
Config & Stage 1 & Stage 2 & $\Delta$ \\
\midrule
PANNs-8 & 66.2 & 66.9 & $+0.7$ \\
Whisper-60s-chunked (Q-proj.\ tuned) & \best{96.0} & 95.5 & $-0.5$ \\
Whisper-60s-chunked (Q-proj.\ frozen) & \best{96.0} & \best{97.3} & $+1.3$ \\
\bottomrule
\end{tabular}
\end{table}

\subsection{Comparison with Qwen2.5-Omni}
\label{sec:omni}

To test whether our advantage is just task fine-tuning, we fine-tune Qwen2.5-Omni-7B under conditions matched
to our Stage 2: the same available-video training and test pairs, the same 8 precomputed
frames, the same precomputed TTS audio (so the omni model must also decode the spoken question), the same
30\,s of music, and the same Stage-2 hyperparameters (1 epoch, effective batch 16, LoRA $r{=}64$ on its
``thinker''). This is a controlled \emph{system-level} comparison, not a confound-free encoder swap: the two
systems share data, inputs, and Stage-2 hyperparameters, but still differ in several ways. Our system
additionally ran a projector-alignment Stage 1 and inherited an ASR-aligned audio projector, whereas Omni
received a single AVQA epoch with no such warmup; the two use different base LLMs (Qwen2-VL vs.\
Qwen2.5-Omni) and different audio pathways (Omni feeds its audio-encoder tokens at their native rate; we feed
a Whisper frame sequence held at a constant 0.94\,s/token); and we report a single Omni seed. The 30\,s music
restriction is a \emph{processor} default in Omni's pipeline, not an architectural limit (both audio encoders
are Whisper-derived), so we make the primary comparison against our \emph{30\,s} model (Whisper-30s), which
encodes the same 30\,s of music, and report the full-duration model (Whisper-60s-chunked, ${\sim}60$\,s)
alongside it (Table~\ref{tab:omni}).

At matched 30\,s audio, Whisper-30s reaches 95.9\% against Omni's 80.9\%, a \best{15.0}-point gap with
duration, data, frames, and question held fixed. We read this as a useful matched baseline rather than a
clean measurement of the audio encoder alone, given the backbone and adaptation differences above. The
deficit is not uniform (Table~\ref{tab:omni}): it is smallest on Audio/Counting
(+8.0), a pure auditory-discrimination task where a general encoder is competitive, and largest on the
Comparative types: AV/Comparative (+26.9) and Audio/Comparative (+19.1), followed by AV/Location (+17.6).
Comparative questions require tracking relative acoustic magnitudes across time. AV/Location questions
require identifying the sounding or loudest instrument and grounding it to a spatial position in the
video. The larger gaps are therefore consistent with local audio features supporting relative-magnitude
tracking and audio-to-visual source grounding; however, the system-level comparison
does not isolate this mechanism from the backbone and adaptation differences, and we do not directly probe
Omni's audio representation. The gap is not confined to audio-grounded
types (Omni also trails on Visual/Counting and Visual/Location, so part of the deficit reflects general task
adaptation), but the largest deficits occur on comparative and audio-visual localization questions that stress
relative magnitude and spatial source grounding. This is consistent with, though not proof of, a representational rather
than a scale effect.

Full-duration audio (Whisper-60s-chunked, ${\sim}60$\,s) widens the overall margin only modestly, to
+16.4. The two largest increases are on Audio/Comparative, where the gap grows from +19.1 to +25.9, and
AV/Temporal, where it grows from +11.3 to +17.3. Only the AV/Temporal increase has a direct
duration-based reading, since ordering events across a 60\,s clip needs audio the 30\,s model never sees
(Section~\ref{sec:why}); the comparative types carry the widest cross-seed spread
(Section~\ref{sec:seeds}), so we do not lean on the Audio/Comparative increase. Temporal reasoning is
thus plausibly a \emph{duration} effect layered on top, separate from the matched-duration comparison.

\begin{table}[t]
\centering
\caption{\textbf{Per-type gap vs.\ fine-tuned Qwen2.5-Omni-7B (test, $n{=}7{,}402$).} Omni's default audio
preprocessing truncates each clip to its first 30\,s, so the matched-duration comparison is our 30\,s model
(Whisper-30s, $\Delta_{30}$); we also show the full-duration model
(Whisper-60s-chunked, ${\sim}60$\,s, $\Delta_{60}$). The systems also differ in backbone and adaptation
(Section~\ref{sec:omni}), so this is a system-level baseline, not an isolated encoder swap. Rows
sorted by $\Delta_{30}$; per-type sample counts are listed in Table~\ref{tab:examples}. All three
systems are evaluated at the shared default seed 42, so each $\Delta$ is a within-seed difference;
the comparative rows carry the largest cross-seed spread (Section~\ref{sec:seeds}), so the matched
$+15.0$ overall gap, not the exact per-type deltas, is the robust quantity.}
\label{tab:omni}
\tablebodyfont
\begin{tabular}{l c c c c c}
\toprule
Question type & Omni & Whisper-30s & $\Delta_{30}$ & 60s-chunked & $\Delta_{60}$ \\
\midrule
Audio / Counting       & 89.4 & 97.4 & $+8.0$  & 97.9 & $+8.5$ \\
AV / Temporal          & 74.9 & 86.2 & $+11.3$ & 92.2 & $+17.3$ \\
Visual / Counting      & 86.9 & 98.6 & $+11.7$ & 99.0 & $+12.1$ \\
AV / Existential       & 85.2 & 98.1 & $+12.9$ & 99.2 & $+14.0$ \\
Visual / Location      & 82.8 & 96.6 & $+13.8$ & 95.6 & $+12.8$ \\
AV / Counting          & 82.6 & 98.5 & $+15.9$ & 98.9 & $+16.3$ \\
AV / Location          & 76.1 & 93.7 & $+17.6$ & 97.6 & $+21.5$ \\
Audio / Comparative    & 69.6 & 88.7 & $+19.1$ & 95.5 & $+25.9$ \\
AV / Comparative       & 69.2 & 96.1 & $+26.9$ & 97.0 & $+27.8$ \\
\midrule
\textbf{Overall}       & 80.9 & 95.9 & \best{$+15.0$} & 97.3 & $+16.4$ \\
\bottomrule
\end{tabular}
\end{table}

\subsection{Dependence on music and question audio}
\label{sec:modality}

We test dependence on the audio inputs by modifying the inputs to already-trained checkpoints at
inference time; no weights are updated. In the \emph{$-$ music} condition, the projected music
embeddings are replaced with zeros immediately before token concatenation, while the video frames,
TTS-spoken question, and fixed prompt are retained. In the \emph{$-$ all audio} condition, the music
embeddings are zeroed and the spoken-question tokens are also removed, leaving only the visual stream
and fixed prompt. Thus, $-$music is the direct test of dependence on music evidence, whereas $-$all
audio is a deliberately degenerate no-question condition and should not be interpreted as the
contribution of music alone.

For Whisper-60s-chunked, removing only the music embeddings reduces overall accuracy from 97.3\%
to 77.2\%, a 20.1-point drop. AV/Temporal falls by 35.5 points, Audio/Comparative by 33.6, and
AV/Location by 32.2, showing that the trained model relies most on the music sequence for event
ordering, relative-magnitude judgments, and cross-modal localization. Because the checkpoint was trained with music present, the 77.2\%
result is not the accuracy of a separately optimized vision-plus-question system.

When the spoken question is removed as well, overall accuracy falls to 53.7\%. The Comparative
categories then rise relative to the $-$music condition: Audio/Comparative increases from 61.9\% to
79.4\%, and AV/Comparative from 77.5\% to 80.3\%. This reversal is not improved comparative
reasoning---without the question, the model cannot know which comparison was asked---and is instead
consistent with frequent-answer or visual-label priors in the no-question setting.

We apply the same full-versus-$-$all-audio comparison to the trained PANNs-32 checkpoint. Its overall
accuracy falls from 69.9\% to 55.8\%, while Audio/Comparative rises from 38.1\% to 78.5\% and
AV/Comparative from 54.0\% to 76.4\%. In both checkpoints, the no-question condition lifts the
Comparative categories into a similar 76--80\% band, supporting the
interpretation that the Comparative anomaly is a property of the degenerate no-question condition
rather than of the Whisper representation. With all inputs present, PANNs-32 is also weakest on the two
Comparative categories among the types shown in Table~\ref{tab:modality}
(38.1\% and 54.0\%, versus 69.9\% overall), whereas Whisper reaches 95.5\% and 97.0\%. This
pattern is consistent with a globally pooled music representation being particularly limited for
relative-magnitude reasoning, although differences in category difficulty and model configuration mean
that it is supporting evidence rather than an isolated causal test.

\begin{table}[H]
\centering
\caption{\textbf{Inference-time input ablations on trained models ($n{=}7{,}402$).} No model is
retrained. For Whisper-60s-chunked, ``$-$ music'' zeros the projected music embeddings while keeping
frames, the TTS-spoken question, and the fixed prompt. For both Whisper-60s-chunked and PANNs-32,
``$-$ all audio'' also removes the spoken-question tokens, leaving vision plus the prompt. The PANNs
rows provide a second-model check on the no-question Comparative reversal
(Section~\ref{sec:modality}).}
\label{tab:modality}
\tablebodyfont
\begin{tabular}{l c c c c c}
\toprule
Configuration & Overall & AV/Temp. & AV/Loc. & Audio/Comp. & AV/Comp. \\
\midrule
Whisper-60s-chunked: full                     & 97.3 & 92.2 & 97.6 & 95.5 & 97.0 \\
\quad $-$ music (question retained)          & 77.2 & 56.7 & 65.4 & 61.9 & 77.5 \\
\quad $-$ all audio (vision + prompt; no Q)  & 53.7 & 44.4 & 40.2 & 79.4 & 80.3 \\
\midrule
PANNs-32: full                                & 69.9 & 69.4 & 65.4 & 38.1 & 54.0 \\
\quad $-$ all audio (vision + prompt; no Q)  & 55.8 & 56.0 & 42.6 & 78.5 & 76.4 \\
\bottomrule
\end{tabular}
\end{table}

\subsection{Residual error distribution and uncertainty}
\label{sec:errors}

The representative model scores 97.3\% on 7{,}402 test questions, leaving approximately 199 errors.
Per-type accuracy identifies where those errors concentrate but does not by itself identify their
cause. AV/Temporal (92.2\%) contributes approximately 50 errors and Visual/Location (95.6\%)
approximately 45; together they account for nearly half of the residual set. These two groups are the
clearest priorities for manual follow-up because they respectively require event ordering over a long
clip and spatial localization from eight sampled frames. We did not manually adjudicate the clips or
predictions, so we do not attribute the errors to chunk boundaries, frame sampling, annotation noise,
or any other specific mechanism.

The high overall accuracy also changes how small differences should be interpreted. The strongest
Whisper variants differ by only one or two points, which is smaller than the 3.9-point cross-seed
standard deviation reported in Section~\ref{sec:seeds}; those orderings are therefore descriptive. The
26.0-point Whisper-30s versus PANNs-32 contrast is much larger and is measured with the same seed,
data, LLM, 32-token budget, and evaluation protocol. However, each encoder ablation was trained once,
so the exact 26.0-point value has no multi-seed uncertainty interval. The supported conclusion is that
preserving fine-grained local temporal detail yields a large accuracy advantage in this protocol, while
the last decimal and small gaps near the ceiling should not be over-interpreted.

\subsection{Robustness to question rephrasing}
\label{sec:robustness}

MUSIC-AVQA questions are generated from a small set of templates, so a model could reach high accuracy
by latching onto template patterns rather than reasoning over the audio and video.
MUSIC-AVQA-R~\citep{ma2024musicavqar} is a benchmark built to detect exactly this: it rephrases each
MUSIC-AVQA test question into many surface forms, split into frequent (\emph{head}) and rare
(\emph{tail}) phrasings, on which template-reliant methods degrade sharply. We use it to test whether
the final model's performance remains robust when the question wording changes, evaluating our best
model (Whisper-60s-chunked) unchanged on a
random $3{,}000$-question sample of each split, drawn from the videos available to us; because our
questions are posed as synthesized speech, this tests rephrased questions delivered through the same
spoken-question interface used during training.

Accuracy is nearly unchanged: 96.5\% on head and 95.6\% on tail, a $0.9$-point
head-to-tail difference; the lower tail score is $1.7$ points below the standard-test 97.3\%. Because the two figures come
from different random question samples and questions are clustered by source video, we treat this as a
descriptive comparison rather than a formal significance test. No type shows a large tail degradation
in these samples (Table~\ref{tab:musicavqar}): the largest decline is $2.4$ points, four of the nine
types improve, and the audio-grounded comparative types
stay in the $92$--$100\%$ range. Compared with the standard-test per-type figures, the largest
single-sample deviations are AV/Temporal ($87.3\%$ head, $91.7\%$ tail, versus $92.2\%$) and
Visual/Location ($91.5\%$ head versus $95.6\%$); these cross-sample comparisons involve different
question sets and type mixes, so we note them without further interpretation.
The final model therefore remains accurate when questions are rephrased,
providing evidence that its accuracy is not solely dependent on the original question templates. (We did not re-run the PANNs or pooled baselines on
MUSIC-AVQA-R, so this speaks to the robustness of the final model, not to whether the Whisper-vs-PANNs gap
itself is preserved under rephrasing.) We report a random subsample rather than the full rephrased split,
which is $\sim$25$\times$ larger. One caveat specific to our
interface: questions are cached as synthesized speech, so TTS duration and token count may correlate with
question templates and were not separately controlled.

\section{Discussion}
\label{sec:discussion}

\paragraph{Local temporal detail and downstream accuracy.}
Across the evaluated configurations, downstream accuracy tracks the preservation of fine-grained local
temporal information. Whisper sequences at ${\sim}0.94$\,s per token reach 95.9--97.3\%, whereas a
representation of up to 60\,s compressed to ${\sim}1.875$\,s per token reaches 70.5\%. Globally pooled
PANNs configurations, which collapse the clip into one vector and expose no explicit local temporal
sequence to the projector, reach 66.9--69.9\%.
The matched 32-token Whisper--PANNs comparison is the limiting case of this broader pattern, and
the one in which duration and projector capacity both favor the losing side (Section~\ref{sec:why}). A
matched fine-tuned omni-modal system also trails our model (Section~\ref{sec:omni}). We frame this as a
graded relationship rather than a categorical one, and we do not claim an established quantitative law
relating seconds per token to accuracy: the encoders differ along several axes at once, a pooled vector
has no well-defined seconds-per-token value once expanded, and each configuration was trained once. We
conjecture, without claiming to have isolated it, that tasks requiring fine-grained reasoning over audio
will similarly benefit from representations that retain local temporal detail over pooled global vectors. This interpretation is consistent with recent work that treats frame-wise event
localization~\citep{wu2025flam} and temporal grounding in music QA~\citep{kwon2026mustbench} as unresolved
capabilities, although those benchmarks do not provide a direct evaluation of our model.

\paragraph{Fusion through pretrained LLM self-attention.}
The LLM's pretrained self-attention fuses three modalities with no added cross-modal machinery, and most of
the accuracy is reached with the LLM frozen (Section~\ref{sec:stages}). In this setting, strong performance
is possible without adding a task-specific fusion module, which motivates careful attention to the input
audio representation before introducing additional fusion machinery.

\paragraph{Relation to concurrent work.}
\citet{you2025musicavqa} argue that Music AVQA demands specialized spatial-temporal designs
beyond what general MLLMs provide. Our results agree in spirit: a general-purpose omni-modal backbone
underperforms by 15.0 points at the matched 30\,s duration; the gap is 16.4 points against our
full-duration variant. Our results show that a strong Music AVQA system can be built without a specialized
spatial-temporal fusion module; they do not establish that explicit fusion would provide no further benefit.

\paragraph{Modular reuse and efficient task adaptation.}
A frozen Whisper encoder with a small projector compares favorably with the fine-tuned Omni baseline
under matched inputs, showing that modular reuse of a specialist encoder can be highly effective for a
focused audio-understanding task. Frozen reuse also keeps the task-specific training small. The
30--60\,s music tracks are encoded once by
Whisper and cached, so AVQA Stage~1 trains only the 4.6M-parameter music projector and takes
approximately 2.5 A100-hours. Stage~2 adds Qwen2-VL LoRA adaptation, bringing the complete two-stage
AVQA training to approximately 5 A100-hours on one A100 80GB. This design adds a high-accuracy music
pathway without updating Whisper, the vision encoder, or Qwen2-VL's base weights. Qwen2.5-Omni also
starts from a Whisper-large-v3-initialized audio encoder, but its reported pretraining trains modality
encoders and later unfreezes all parameters for broader multimodal learning~\citep{qwen2025omni}. Our
reported hours are marginal AVQA adaptation costs; they exclude the inherited Whisper and Qwen2-VL
pretraining and the reusable ASR initialization in Section~\ref{sec:asr}.

\section{Limitations}
\label{sec:limitations}

\paragraph{Single-task fine-tuning degrades ASR.}
Our AVQA models are not simultaneously good at transcription. Adding the three music special tokens
(Section~\ref{sec:music}) with random embeddings at AVQA Stage 1 already lifts LibriSpeech WER from
5.09\% to 13.57\% (both $n{=}500$, normalized; the full-split baseline is 4.85\%,
Section~\ref{sec:asr})~\emph{even with the LLM and audio projector frozen}: the random
embeddings are occasionally emitted during greedy decoding and corrupt transcripts. Stage 2 LoRA re-aligns
the LLM around the new vocabulary and partially recovers WER to 10.96\%. The damage traces to the newly added vocabulary,
not to touching the audio projector. Freezing the projector (our best AVQA model) is not the main
WER lever. Recovering both capabilities would require multi-task training or task-specific adapter
switching.

\paragraph{Question delivery.}
Our pipeline poses questions as synthesized speech through the Whisper encoder (Section~\ref{sec:tts});
we also probed a variant that delivers the question as native text tokens. Under our projector-only
alignment recipe it trains poorly (67--72\% across configurations), and control experiments attribute
the deficit to a degraded Stage-1 alignment (the projector-only stage fails to converge usefully when
the question bypasses the audio pathway) rather than to an inherent property of text delivery. We
therefore do not read this as evidence for or against either interface; isolating the effect of the
delivery pathway would require an alignment recipe that succeeds under both.

\paragraph{Music-specialist encoders.}
We did not evaluate any music-domain encoder, such as
MERT~\citep{li2023mert}. Because music features are cached offline, such a
comparison was not fundamentally precluded by training memory; its main cost
would be an additional feature-extraction experiment and a second audio
backbone at deployment, alongside Whisper for the spoken question. We instead
prioritized the shared-encoder design, which reuses one pretrained feature
family and already reaches 97.3\% accuracy. Given the limited remaining aggregate-accuracy
headroom and the observed seed variation, establishing a smaller incremental
gain would require a multi-seed MERT study. We therefore leave open whether
music-specific pretraining improves accuracy or stability beyond the
shared-Whisper configuration.

\paragraph{Other limitations.}
\begin{itemize}
\setlength{\itemsep}{2pt}
\item \textbf{Offline encoder.} Music features are precomputed, so inference on a new video needs a
separate encoding pass before the LLM runs.
\item \textbf{Closed answer set.} The 42-answer vocabulary makes exact match a natural metric but does
not measure open-ended generation.
\item \textbf{Single benchmark and backbone.} All results are on MUSIC-AVQA with a
Qwen2-VL-7B-Instruct backbone; whether the local-versus-pooled pattern transfers to other audio-visual tasks or LLMs is
untested.
\item \textbf{Entangled ASR pretraining.} Every AVQA model inherits the ASR-aligned audio projector,
so we cannot fully separate the contribution of ASR pretraining from AVQA training.
\end{itemize}

\section{Conclusion}

Local Whisper frame features provide a lightweight, high-accuracy route to music audio-visual question
answering. By reusing one frozen Whisper encoder for both music and spoken questions, aligning the new
music path with a 4.6M-parameter linear projector, and relying on Qwen2-VL's pretrained self-attention
for fusion, Qwen-MusicAVQA-7B reaches 97.3\% on the available-video MUSIC-AVQA test subset
(96.0\%$\,\pm\,$3.9\% across three full retraining seeds) on a single A100. Most of that accuracy is
already present after projector-only alignment, before LLM LoRA adaptation.

Across the evaluated configurations, the clearest empirical pattern is representational: accuracy tracks how much
fine-grained local temporal information the audio representation preserves. Whisper frames at
${\sim}0.94$\,s per token substantially outperform both a coarser ${\sim}1.875$\,s-per-token
compression of up to 60\,s and globally pooled PANNs embeddings, including a 26-point gain over PANNs at
the same 32-token budget. The modular system also exceeds a matched-duration,
fine-tuned Qwen2.5-Omni baseline by 15 points, while avoiding audio-encoder and base-model retraining for
the task. These results do not compare total foundation-model pretraining costs---our system inherits both
Whisper and Qwen2-VL---but they show that strong task-specific audio capability can be added with a small
learned interface rather than relearning audio perception or introducing a specialized fusion network.
Code is available on
\href{https://github.com/MKDehdashti/Qwen2-vl-audio}{GitHub} and checkpoints in our
\href{https://huggingface.co/MayaKD/qwen2-vl-audio}{Hugging Face repository}.

\FloatBarrier
\bibliographystyle{plainnat}
\bibliography{references}

\clearpage
\appendix
\section*{\LARGE Appendix}
\renewcommand{\thetable}{A\arabic{table}}
\setcounter{table}{0}
\makeatletter
\renewcommand\section{\@startsection{section}{1}{\z@}%
  {-3.0ex \@plus -0.8ex \@minus -.2ex}%
  {1.8ex \@plus .2ex}%
  {\normalfont\large\bfseries}}
\makeatother
\section{MUSIC-AVQA Question Types and Examples}
\label{app:examples}

MUSIC-AVQA organizes questions by the evidence required and the reasoning operation. Audio questions
can be answered from acoustic evidence, Visual questions from visible content, and Audio-Visual (AV)
questions require relating what is heard to what is seen. The reasoning labels cover presence
(Existential), number (Counting), spatial position (Location), relative attributes such as louder or more
rhythmic (Comparative), and event order (Temporal). Not every modality--reasoning combination appears,
which yields the nine official types used in the per-type analyses in
Tables~\ref{tab:omni},~\ref{tab:modality}, and~\ref{tab:musicavqar}. Table~\ref{tab:examples} gives one representative test
question for each type and illustrates the closed, template-derived answer format.

\begin{table}[H]
\centering
\caption{\textbf{One example test question per MUSIC-AVQA question type, with per-type sample counts.}
$n$ is the number of test questions of each type in our available-video subset (total $7{,}402$); these
are the denominators behind the modality columns of Table~\ref{tab:main}, every per-type figure in
Tables~\ref{tab:omni} and~\ref{tab:modality}, and the error decomposition in Section~\ref{sec:errors}. Questions are shown verbatim as rendered
from the dataset's templates, including their grammatical quirks (e.g.\ ``leftest''), because this is
exactly the text our pipeline synthesizes as speech (Section~\ref{sec:tts}); answers come from the
fixed 42-answer vocabulary.}
\label{tab:examples}
\tablebodyfont
\renewcommand{\arraystretch}{1.12}
\begin{tabular}{@{}l c p{0.50\textwidth} l@{}}
\toprule
Question type & $n$ & Example question & Answer \\
\midrule
Audio / Counting     & 894  & How many musical instruments were heard throughout the video? & one \\
Audio / Comparative  & 247  & Is the piano louder than the violin? & no \\
Visual / Counting$^\ast$ & 1{,}072 & Are there accordion and violin instruments in the video? & yes \\
Visual / Location    & 1{,}024 & What kind of instrument is the leftest instrument? & violin \\
AV / Existential     & 894  & Is the violin in the video always playing? & no \\
AV / Counting        & 949  & How many instruments are sounding in the video? & two \\
AV / Location        & 712  & Where is the loudest instrument? & left \\
AV / Comparative     & 972  & Is the instrument on the right more rhythmic than the instrument on the left? & yes \\
AV / Temporal        & 638  & What is the first instrument that comes in? & piano \\
\midrule
\textbf{Total}       & \textbf{7{,}402} & & \\
\bottomrule
\end{tabular}

\vspace{2pt}
\raggedright\tablenotefont
$^\ast$ Visual/Counting is the benchmark's official label for this template, even though the example's
yes/no form reads as existential. We reproduce the dataset's taxonomy unchanged.
\end{table}

\section{Per-type accuracy on MUSIC-AVQA-R}
\label{app:musicavqar}

\begin{table}[H]
\centering
\caption{\textbf{Per-type accuracy on MUSIC-AVQA-R (Whisper-60s-chunked, unchanged).} Head
(frequent) and tail (rare) phrasings, each a random $3{,}000$-question sample drawn from the videos
available to us. The benchmark's diagnostic for template reliance is the head-to-tail change:
no type loses more than $2.4$ points, and four types improve. The two samples differ in type
composition (e.g.\ Visual/Counting is $18.1\%$ of head but $4.7\%$ of tail), so the per-type rows
provide the more comparable head-to-tail diagnostic. A simple reweighting of the reported per-type
accuracies to the opposite split's type distribution reduces the $0.9$-point aggregate gap to
approximately $0.3$--$0.6$ points. Because head and tail are unpaired random samples, the per-type
differences are descriptive; no significance tests are reported.
$\Delta$ is computed from unrounded accuracies and may therefore
differ by $0.1$ from the difference of the rounded columns.}
\label{tab:musicavqar}
\tablebodyfont
\begin{tabular}{l c c c c c}
\toprule
 & \multicolumn{2}{c}{Head} & \multicolumn{2}{c}{Tail} & \\
\cmidrule(lr){2-3}\cmidrule(lr){4-5}
Question type & Acc.\ (\%) & $n$ & Acc.\ (\%) & $n$ & $\Delta$ \\
\midrule
Audio / Counting      & 96.8  & 463 & 95.6  & 114 & $-1.2$ \\
Audio / Comparative   & 92.5  & 93  & 100.0 & 125 & $+7.5$ \\
Visual / Counting     & 100.0 & 543 & 97.9  & 141 & $-2.1$ \\
Visual / Location     & 91.5  & 400 & 92.2  & 678 & $+0.7$ \\
AV / Existential      & 100.0 & 286 & 98.0  & 540 & $-2.0$ \\
AV / Counting         & 99.7  & 365 & 97.4  & 378 & $-2.4$ \\
AV / Location         & 93.6  & 326 & 93.8  & 194 & $+0.3$ \\
AV / Comparative      & 100.0 & 304 & 97.7  & 517 & $-2.3$ \\
AV / Temporal         & 87.3  & 220 & 91.7  & 313 & $+4.4$ \\
\midrule
\textbf{Overall}      & \textbf{96.5} & \textbf{3{,}000} & \textbf{95.6} & \textbf{3{,}000} & $-0.9$ \\
\bottomrule
\end{tabular}
\end{table}

\end{document}